\documentclass{iau}
\usepackage{graphicx}

\title[Environments of radio-loud AGN] %% [give here short title] %%
{The cluster environments of radio-loud AGN}

\author[J Ineson et al.]   %% [give here the short author list; use "et al." if 3 authors or more] %%
{
Judith Ineson$^1$, Judith Croston$^1$, Martin Hardcastle$^2$, Ralph Kraft$^3$, Daniel Evans$^3$, Matt Jarvis$^{4,5}$
}

\affiliation{
$^1$School of Physics and Astronomy, University of Southampton,
  Southampton SO17 1BJ, UK \\ 
email: {\tt ji1g10@soton.ac.uk} \\
$^2$School of Physics, Astronomy and Mathematics, University of
  Hertfordshire, Hatfield, Hertfordshire AL10 9AB, UK \\
$^3$Harvard-Smithsonian Center for Astrophysics, 60 Garden Street, Cambridge, MA 02138, USA\\
$^4$Astrophysics, Department of Physics, Keble Road, Oxford OX1 3RH, UK\\
$^5$Department of Physics, University of the Western Cape, Private Bag
X17, Bellville 7535, South Africa\\
}

\pubyear{2014}
\volume{313} 
\pagerange{xx--xx}
\jname{Extragalactic jets from every angle}
\editors{F. Massaro, C.C. Cheung, E. Lopez, A. Siemiginowska, eds.}
\begin{document}

\maketitle

\begin{abstract}

%\keywords{Keyword1, keyword2, keyword3, etc.}
Radio-loud AGN play an important r\^ole in galaxy evolution. We need to understand their properties, and the processes that affect their behaviour in order to model galaxy formation and development. We here present preliminary results of an investigation into the cluster environments of radio galaxies. We have found evidence of a strong correlation between radio luminosity and environment richness for low excitation radio galaxies, and no evidence of evolution of the environment with redshift. Conversely, for high excitation radio galaxies, we found no correlation with environment richness, and tentative evidence of evolution of the cluster environment.
\end{abstract}

\firstsection % if your document starts with a section,

\section{The ERA programme}

There is a complex, two-way relationship between radio-loud AGN and their cluster environments, and understanding this relationship is crucial for our understanding of the r\^ole of AGN feedback in galaxy evolution (eg \cite{McN12}; \cite{Hec14}). The radio jets transport energy a considerable distance into the cluster, and are themselves affected by the intra-cluster medium. Do the properties of the large-scale cluster environment in their turn affect the feedback loop maintaining the AGN, or are the AGN properties solely determined by the more local environment of the host galaxy? And how does does this disruption of the cluster environment affect its evolution?

In the ERA programme (\cite{Ine13}), we are making a systematic examination of the effects of epoch and environment on the properties of radio-loud AGN in order to address two questions: is radio luminosity is related to the large-scale cluster environment, and does the environment evolve with epoch?

In phase 1, we used the ERA sample (Fig.\ref{fig:All}, top left) to compare radio luminosity and the hot gas environment within a limited redshift range ($0.4<z<0.6$), thus removing the effects of redshift evolution. We also looked for differences between the populations of High and Low Excitation Radio Galaxies (HERGs and LERGs), which have different accretion rates and host galaxy properties. We now are comparing these results with a sample at $z<0.2$ to look for evolution effects. We use the X-ray luminosity of the ICM as a proxy for total cluster mass.

\section{Correlation between radio luminosity and cluster environment}

The results from the ERA and $z<0.2$ samples individually show an overall correlation between radio luminosity ($L_{\rm R}$) and cluster luminosity ($L_{\rm X}$), with the correlation is driven entirely by the LERG sub-sample. When the samples are combined (Fig.\ref{fig:All}, top), the $L_{\rm R}-L_{\rm X}$ correlation for the LERGs is very strong ($p<0.00005$, Kendall's $\tau$ test with partial correlation for redshift), and again there is no correlation for the HERGs.

Thus there is strong evidence of a correlation between radio luminosity and the cluster environment for LERGs but not for HERGs, supporting the theory that LERGs take part in a feedback loop connected to the cluster gas.

\section{Investigating redshift evolution}

\begin{figure}[!hbp]
% \vspace*{-2.0 cm}
\begin{center}
 \includegraphics[width=5in]{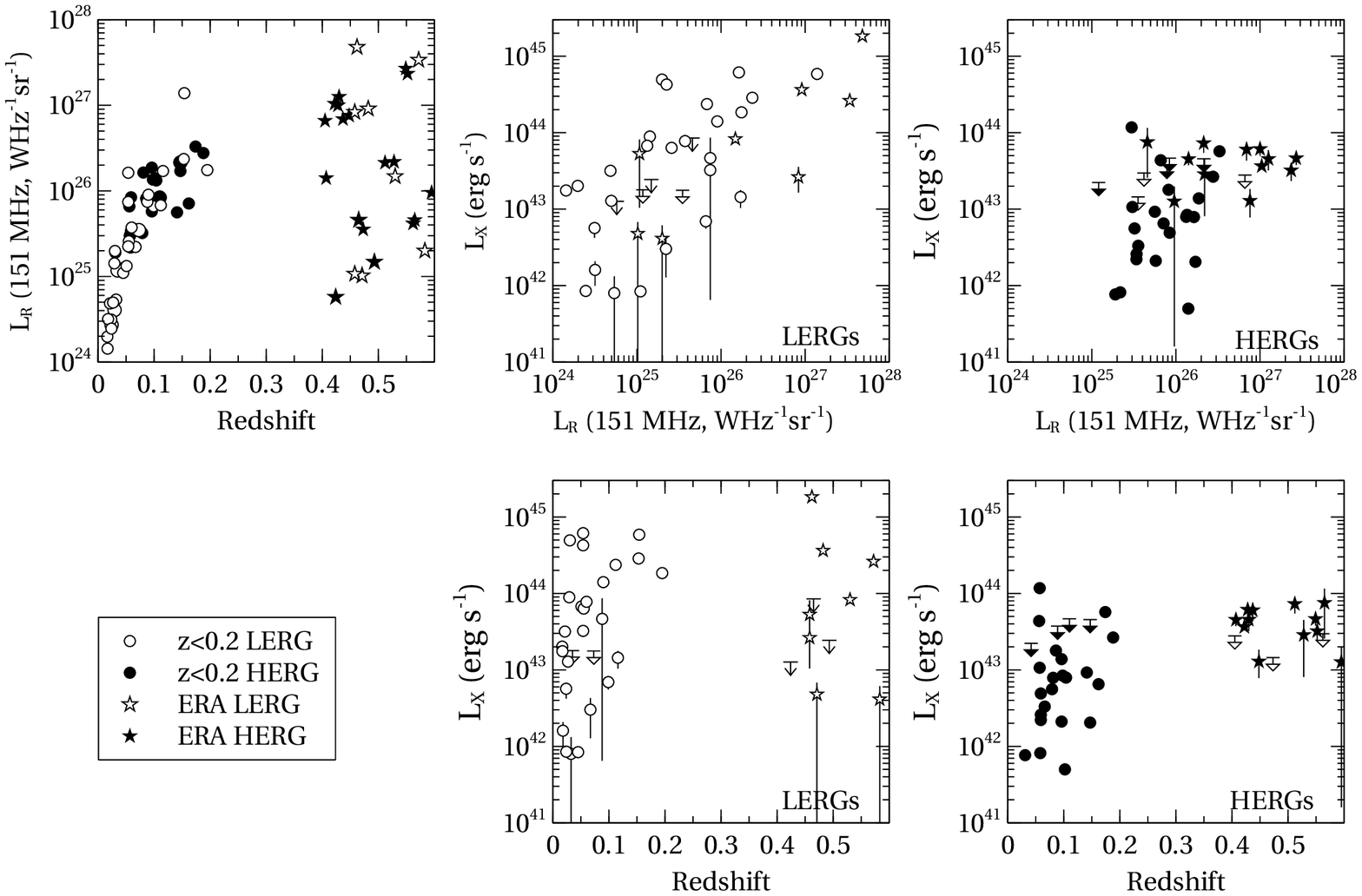} 
% \vspace*{-1.0 cm}
 \caption{On the left, the $z<0.2$ and ERA samples. Results for the combined samples are shown in the centre and right, with radio luminosity vs cluster richness (top) and redshift vs cluster richness for sub-samples matched in radio luminosity (bottom). LERGs are in the central column, and HERGs on the right.}
   \label{fig:All}
\end{center}
\end{figure}

Both LERG sub-samples occupy the same region of the $L_{\rm R}-L_{\rm X}$ plot, and sub-samples matched in radio luminosity (Fig.\ref{fig:All}, bottom) have comparable distributions of $L_{\rm X}$, implying that they come from the same population. Conversely, the $z<0.2$ HERGs occupy a wider range of environments than ERA HERGs, and the matched sub-samples suggest that they come from different populations ($p<0.01$, Peto \& Prentice tests). The ERA subsamples are however small and contain upper limits.

This gives tentative evidence of evolution of HERG but not LERG environments, again suggesting a difference in the feedback r\^oles of the two classes (\cite{Hec14}).

\end{document}